\documentclass[useAMS,usegraphicx,usenatbib,usedcolumn,preprint]{aastex}
\usepackage{spr-astr-addons}
\usepackage{lscape}
\usepackage{latexsym}
\usepackage{graphics}
\usepackage{sidecap}
\usepackage{subfigure}
\usepackage{textcomp}
\usepackage{bbding}
\usepackage{epsf}
\usepackage{amsmath,amssymb}
\usepackage{sidecap}
\usepackage{multirow}
\usepackage{flafter}
\usepackage{float}
\usepackage{color}
\usepackage{longtable}
\usepackage{rotating}
\makeatletter

\RequirePackage{color}

\newcommand{\NGC}{NGC\,300}

\newcommand{\HII}{\hbox{H\,{\sc ii}}}
\newcommand{\HI}{\hbox{H\,{\sc i}}}

\newcommand{\BKG}{BKG}

\newcommand{\SNR}{SNR}
\newcommand{\xrb}{xrb}

\def\p0{\phantom{0}}

\begin{document}
\title{Radio-Continuum study of the Nearby Sculptor Group Galaxies. Part~1: \NGC\ at $\lambda$=20~cm}
 \author{Timothy J. Galvin, Miroslav D. Filipovi\'c, Evan J. Crawford, Graeme Wong, Jeff L. Payne, Ain De Horta, Graeme L. White\altaffilmark{1}, Nick Tothill, Danica Dra\v skovi\'c} 
  \affil{University of Western Sydney, Locked Bag 1797, Penrith South DC, NSW 1797, Australia}
\and 
 \author{Thomas G. Pannuti, Caleb K. Grimes, Benjamin J. Cahall}
  \affil{Department of Earth and Space Sciences, Space Science Center, 235 Martindale Drive, Morehead State University, Morehead, KY 40351, USA}

 \author{William C. Millar\altaffilmark{2}}
  \affil{Centre for Astronomy, James Cook University, Townsville, Queensland 4811, Australia}
   \email{wmillar@grcc.edu}
\and
 \author{Seppo Laine}
  \affil{Spitzer Science Center, California Institute of Technology, MS 220-6, Pasadena, CA 91125, USA}
\and
\altaffiltext{1}{Charles Sturt University, School of Dentistry, Wagga Wagga NSW, Australia}
\altaffiltext{2}{Grand Rapids Community College, 143 Bostwick N.E., Grand Rapids, MI, 49503, USA}

\begin{abstract}
A series of new radio-continuum ($\lambda$=20~cm) mosaic images focused on the \NGC\ galactic system were produced using archived observational data from the VLA and/or ATCA. These new images are both very sensitive (rms=60~$\mu$Jy) and feature high angular resolution ($<$10\arcsec). The most prominent new feature is the galaxy's extended radio-continuum emission, which does not match its optical appearance. Using these newly created images a number of previously unidentified discrete sources have been discovered. Furthermore, we demonstrate that a joint deconvolution approach to imaging this complete data-set is inferior when compared to an \textsc{immerge} approach.

\end{abstract}

\section{Introduction}

At $\sim$1.9~Mpc away \citep{2006ApJ...638..766R}, \NGC{} is the closest spiral galaxy of the nearby Sculptor Group. This proximity is an advantage because it allows for this galaxy to be examined in great detail. Previous radio-continuum and optical studies of \NGC{} \citep{2000ApJ...544..780P,2004A&A...425..443P,2011Ap&SS.332..221M} utilised either the Australia Telescope Compact Array (ATCA) or the Very Large Array (VLA) as their primary instrument. However, these past studies suffer from either low resolution, poor sensitivity or both.

   
Until the next generation radio telescopes, such as the Australian Square Kilometre Array Pathfinder (ASKAP), Karoo Array Telescope (KAT \& MeerKAT) and Square Kilometre Array (SKA), become operational we are restricted to consolidating a selection of \NGC{} radio observations. In this paper, we reexamine all available archived radio-continuum observations performed at ATCA and VLA at $\lambda$=20~cm ($\nu$=1.4~GHz) with the intention of merging these observations into a single radio-continuum image. By combining a large amount of existing data using the latest generation of computer power we can create new images that feature both high angular resolution and excellent sensitivity. The newly constructed images are analysed and the difference between the various \NGC{} images created at 20~cm are discussed.

In \S2 we describe the observational data and reduction techniques. In \S3 we present our new maps, a brief discussion is given in \S4, and \S5 is the conclusion.

\section{OBSERVATIONAL DATA}
\subsection{Observational Data}

To create a high-resolution and sensitive radio-continuum image a number (5) of observations from the ATCA and VLA were used. These observations were selected from the Australian Telescope Online Archive (ATOA) and the National Radio Astronomy Observatory (NRAO) Science Data Archive. The observations which were selected and used here are summarised in Table~\ref{table:NGCobs}. Because \NGC\  is far south, the VLA could only record data for short durations (see Table~\ref{table:NGCobs}) and therefore over limited hour angle ranges.

ATCA project C1757 was conducted in mosaic mode with a total of eight pointings being observed over eight days. This project mapped the neutral hydrogen emission of \NGC{} \citep{2011MNRAS.410.2217W}, which extends significantly beyond the galaxy's optical emission. We used only the four innermost pointings which were directly centred on \NGC{}. Other VLA and ATCA observations used in this study consist of single pointings of \NGC. All images are primary beam corrected.

\begin{table*}
\tiny
\center
\caption[List of VLA and ATCA observations of \NGC{} used in this study.]{List of VLA and ATCA observations of \NGC{} used in this study. RA and DEC represent coordinates of central pointings.\label{table:NGCobs}}
\begin{tabular}{@{}c  c  c  c c c c c c c c@{} }
\hline
$Project$ & $RA$ $(J2000)$  & $Dec$ $(J2000)$ & $Dates$ & $Instrument$ & $Array$ & $\nu$ & $\Delta\nu$ & $Duration$ & $Primary$ & $Secondary$\\
$Code$&h   \space m \space s  \space\space\space &\space \textdegree   \space\space\space '  \space " \space\space &  &  &$Type$&$(MHz)$&$(MHz)$&$(hours)$&$Calibrator$ & $Calibrator$\\  \hline
AL445   & 00 54 53.20  & -37 40 57.00 & 13/06/1998 & VLA & AB  &1435, 1465&15, 15&1& 0134+329 & 0022-423 \\
AC101   & 00 54 52.72 & -37 41 09.02 & 13/07/1984 & VLA & CD &1465, 1515&50, 50&0.1& 0134+329 & 0023-263 \\ 
AC308   & 00 55 30.00 & -37 51 54.00 & 09/10/1993 & VLA & CD &1365, 1435&50, 50&0.01& 0521+166 & 0116-208\\ 
\smallskip
C828 & 00 54 53.47 & -37 41 00.00 & 27-28/02/2000 & ATCA & 6A &1384&128&10.50& 1934-638 & 0823-500\\
C1757   & 00 54 11.81 & -37 32 49.00 & 20/11/2007 & ATCA & EW367 &1384&128&1.04& 1934-638 & 0008-421 \\
C1757   & 00 54 11.81 & -37 32 49.00 & 02/02/2008 & ATCA & EW367 &1384&128&1.15& 1934-638 & 0008-421  \\
C1757   & 00 55 35.19 & -37 49 19.00 & 03/02/2008 & ATCA & EW367 &1384&128&1.14& 1934-638 & 0008-421  \\
C1757   & 00 55 35.19 & -37 32 49.00 & 23/05/2008 & ATCA & EW367 &1384&128&1.22& 1934-638 & 0008-421  \\
C1757   & 00 54 11.81 & -37 49 19.00 & 24/05/2008 & ATCA & EW367 &1384&128&1.33& 1934-638 & 0008-421  \\
C1757   & 00 55 35.19 & -37 49 19.00 & 13/11/2008 & ATCA & EW367 &1384&128&1.17& 1934-638 & 0008-421  \\
C1757   & 00 55 35.19 & -37 32 49.00 & 17/11/2008 & ATCA & EW367 &1384&128&1.28& 1934-638 & 0008-421  \\
C1757   & 00 54 11.81 & -37 49 19.00 & 19/11/2008 & ATCA & EW367 &1384&128&1.23& 1934-638 & 0008-421 \\

\hline
\end{tabular}
\end{table*}

For both ATCA projects, the source PKS 1934-638 was used as the primary calibrator. The sources PKS 0008-421 and PKS 0823-500 were used as secondary calibrators for ATCA projects C1757 and C828 respectively. VLA observations AC101 and AL445 used source 0134+329 (IAU J0137+3309) as their primary calibrator, while sources 0023-263 (IAU J0025-2602) and 0022-423 (IAU J0024-4202) were used as the secondary calibrator, respectively. Project AC308 used 0521+166 (PKS J0521+1638) as its primary calibrator and source 0116-208 (PKS J0116-2052) as its secondary calibrator.

\begin{figure*}
 	\center{\includegraphics[width=0.5\textwidth, angle=270,trim= 140 180 0 180]{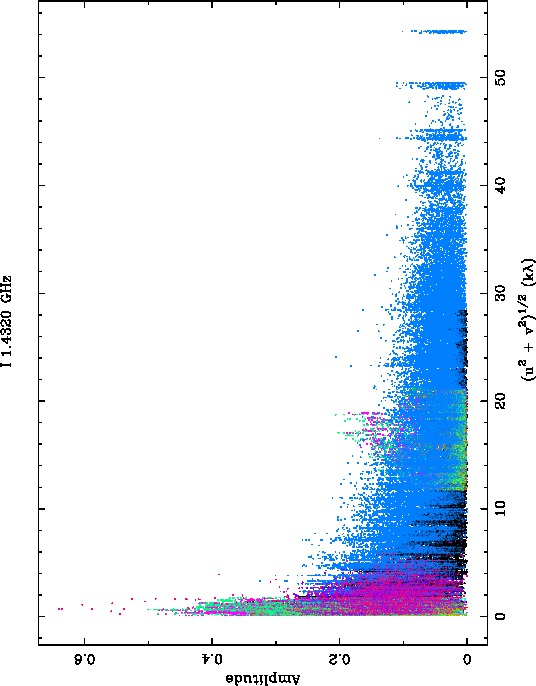}}
	\caption{Graph showing the unrestricted amplitude vs \emph{uv}-distance of all ATCA/VLA projects used in this work. Each project is represented as different colour in this graph. Here, project C828 is dark blue, project C1757 is green, project AL445 is light blue, project AC308 is pink and project AC101 is purple.}
    \label{fig:uvd} 
\end{figure*}

\subsection{Image Creation}

To create the best possible \NGC{} mosaic image at 20~cm, we examined the radial distance in the {\it uv}-plane of all observations combined (Fig.~\ref{fig:uvd}) and found that the {\it uv}-plane is densely sampled up to $\sim$30~k$\lambda$, but quite sparsely sampled at longer baselines. Thus we restricted our imaging to the 0-30~k$\lambda$ range and discarded longer baselines in VLA observations. 

The \textsc{miriad} \citep{miriad}, \textsc{aips} \citep{aips} and \textsc{karma} \citep{karma} software packages were used for data reduction and analysis. Because of the large volume of data, the \textsc{miriad} package was compiled to run on a 16-processor high-performance computer system. 

Initially, observations which were performed using the VLA were imported into \textsc{aips} using the task \textsc{fillm}, and then all sources were split with \textsc{split}. Using the task \textsc{uvfix}, source coordinates were converted from the B1950 to the J2000 reference frame and the task \textsc{fittp} was used to export each source to a \textsc{fits} file. 

The \textsc{miriad} package was then used for actual data reduction. The task \textsc{atlod} was used to convert ATCA observations into \textsc{miriad} files, while the task \textsc{fits} was used to import the previous \textsc{aips}-produced fits files and convert them to \textsc{miriad} files. 

Typical calibration and flagging procedures were then carried out \citep{miriad}. Using the task \textsc{invert}, two mosaic images (one based on ATCA data and the other on VLA data) were created using a natural weighting scheme. Each mosaic image was then cleaned using the task \textsc{mossdi}. The \textsc{mossdi} task is a SDI clean algorithm designed for mosaic images \citep{1984A&A...137..159S}. To convolve a clean model the task \textsc{restor} was then used on each of the cleaned maps. 

The restored mosaic images were then merged together using the task \textsc{immerge}. This task uses a linear approach to merge images of a different resolution, where the low resolution image is assumed to better represent short spacing, and the higher resolution image to better represent the fine structure.  \textsc{immerge} changes the weighting by the normalisation process in the overlapping region. 

We also created a single mosaic image which was comprised of all selected ATCA and VLA projects using the task \textsc{invert}. The dirty map was then cleaned with \textsc{mossdi} before being convolved with \textsc{restor}.   

The difference between the two approaches to data combination is that the single image (in which all {\it uv}-data are inverted simultaneously) contains all calibration uncertainties and weightings from each data set, while the image produced by merging individual images normalises the different calibrations in regions of {\it uv} overlap.

The ATCA bandpass comprises 13 channels, one of which is contaminated by \HI\ emission. From the \HI\ results of \citet{2011MNRAS.410.2217W}, the \HI\ contamination can be seen to be 
restricted to channel 4, which was flagged out and does not contribute to our images.

 \section{RESULTS}
 
\begin{figure*}
 	\center{\includegraphics[width=0.5\textwidth,trim=0 90 0 0,angle=270]{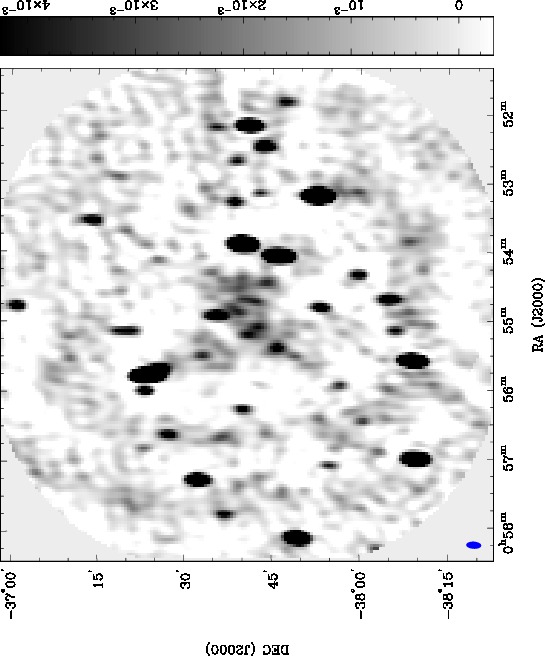}}
	\caption[Mosaic ATCA Project C1757 total intensity image]{ATCA Project C1757 mosaic radio-continuum total intensity image of \NGC. The synthesised beam, as represented by the blue ellipse in the lower left hand corner, is $153.1\arcsec \times 72.5$\arcsec\ and the r.m.s. noise is 0.30~mJy/beam. This image is in terms of Jy/Beam. \label{fig:C1757}}
\end{figure*}

\begin{figure*}
	\center{\includegraphics[width=0.5\textwidth,trim=0 90 0 0,angle=270]{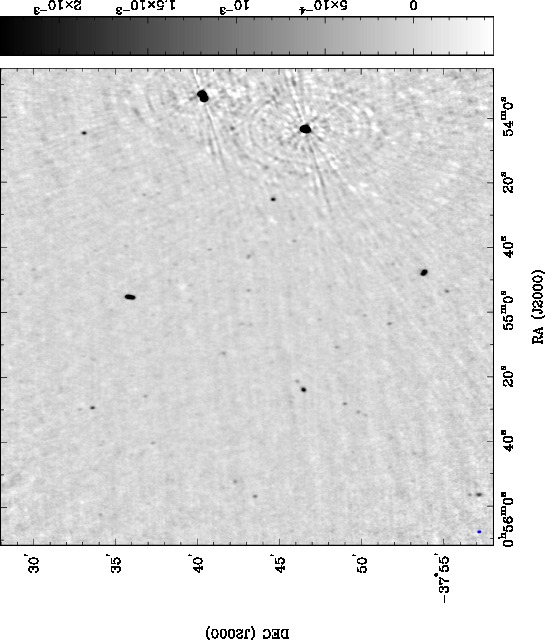}}
	\caption[ATCA Project C828 total intensity image]{ATCA Project C828 radio-continuum total intensity image of \NGC. The synthesised beam, as represented by the blue ellipse in the lower left hand corner, is $13.8\arcsec \times 10.1$\arcsec\ and the r.m.s noise is 0.07~mJy/beam. This image is in terms of Jy/Beam.  \label{fig:C828}}
\end{figure*}

When comparing individual maps created from each observation, the different effects of varying array configurations can be seen. The mosaic image produced from C1757 (Fig.~\ref{fig:C1757}) showed a much better defined region of extended emission. However, it lacked the high resolution of project C828 (Fig.~\ref{fig:C828}), due to the shorter baselines of the C1757 project.

\begin{figure*}
	\center{\includegraphics[width=0.5\textwidth,trim=0 90 0 0,angle=270]{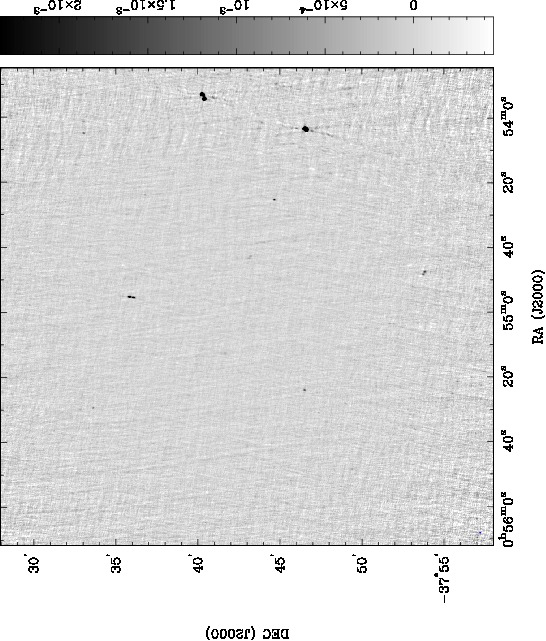}}
	\caption[VLA Project AL0445 total intensity image]{VLA Project AL0445 radio-continuum total intensity image of \NGC. The synthesised beam, as represented by the blue ellipse in the lower left hand corner, is $7.0\arcsec \times 4.5$\arcsec\ and the r.m.s noise is 0.15~mJy/beam. This image is in terms of Jy/Beam. \label{fig:AL445}}
\end{figure*}

\begin{figure*}
 	\center{\includegraphics[width=0.5\textwidth,trim=0 90 0 0,angle=270]{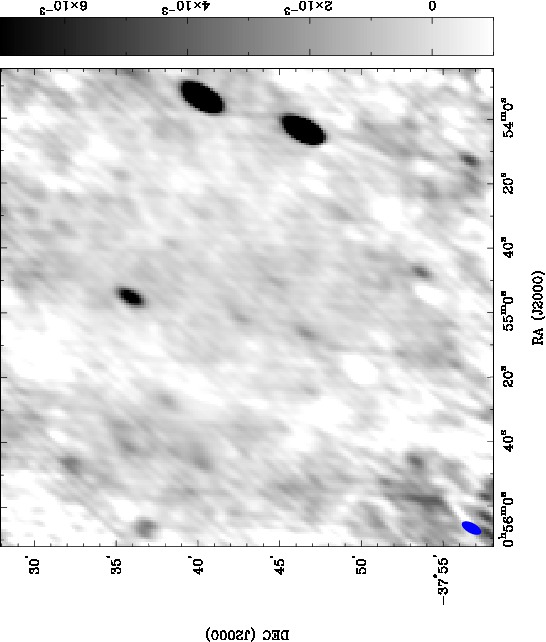}}
	\caption[VLA Project AC0101 total intensity image]{VLA Project AC0101 radio-continuum total intensity image of \NGC. The synthesised beam, as represented by the blue ellipse in the lower left hand corner, is $77.9\arcsec \times  36.7$\arcsec\ and the r.m.s noise is 0.30~mJy/beam.  This image is in terms of Jy/Beam.  \label{fig:AC101}}
\end{figure*}

\begin{figure*}
	\center{\includegraphics[width=0.5\textwidth,trim=0 90 0 0,angle=270]{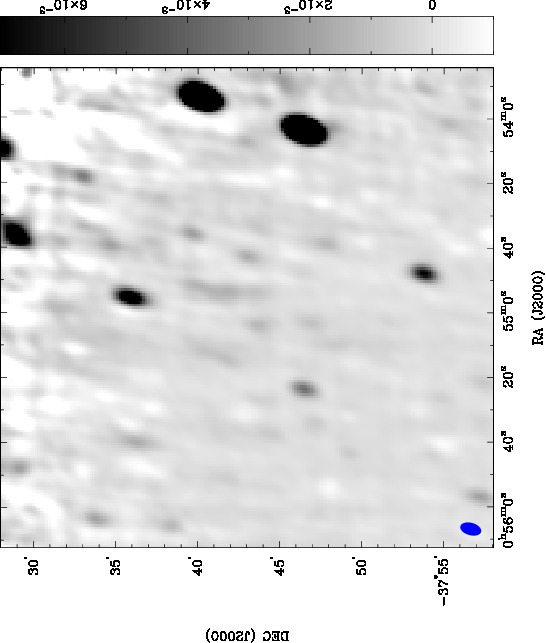}}
	\caption[VLA Project AC0308 total intensity image]{VLA Project AC0308 radio-continuum total intensity image of \NGC. The synthesised beam, as represented by the blue ellipse in the lower left hand corner, is $77.8\arcsec \times 44.7$\arcsec\ and the r.m.s. noise is 0.50~mJy/beam. This image is in terms of Jy/Beam.  \label{fig:AC308}}
\end{figure*}

VLA observation AL445 (Fig.~\ref{fig:AL445}) was performed using the longest baseline, giving this image the highest resolution among all the individual images presented in this paper. As expected, point sources dominate this field. The other VLA observations, AC101 (Fig.~\ref{fig:AC101}) and AC308 (Fig.~\ref{fig:AC308}), were taken with much smaller array configurations, resulting in lower-resolution images with significant extended emission. Both of these observations have short integration times. AC308 had only 36 seconds integration, leading to a very shallow image.  

\begin{figure*}
	\center{\includegraphics[width=0.5\textwidth,trim=0 90 0 0,angle=270]{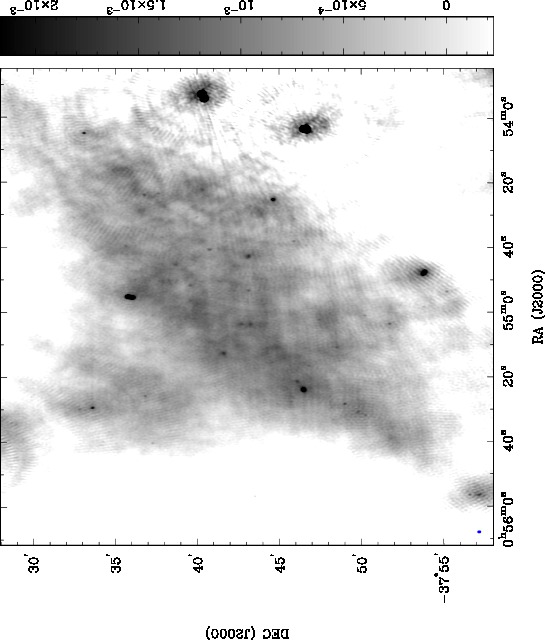}}
	\caption[Combined ATCA Projects C1757 and C828]{Combined ATCA projects C1757 and C828 mosaic radio-continuum total intensity image of \NGC. The synthesised beam, as represented by the blue ellipse in the lower left hand corner, is $13.8\arcsec \times 10.1$\arcsec\ and the r.m.s noise is 0.07~mJy/beam. This image is in terms of Jy/Beam.  \label{fig:atca-mosaic}}
\end{figure*}

\begin{figure*}
	\center{\includegraphics[width=0.5\textwidth,trim=0 90 0 0,angle=270]{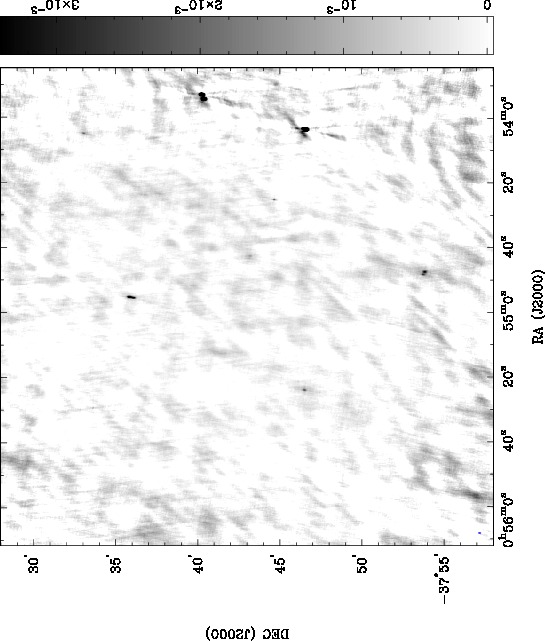}}
	\caption[Combined VLA Projects AL0445, AC0101 and AC0308]{Combined VLA projects AL445, AC101 and AC308 mosaic radio-continuum total intensity image of \NGC. The synthesised beam, as represented by the blue ellipse in the lower left hand corner, is $10.9\arcsec \times 4.8$\arcsec\ and the r.m.s. noise is 0.09~mJy/beam. This image is in terms of Jy/Beam.  \label{fig:vla-all}}
\end{figure*}

Fig.~\ref{fig:atca-mosaic} is a mosaic image containing both ATCA observations C1757 and C828. It shows a well-defined region of extended emission with a number of definitive point sources. We also created a mosaic image comprised of VLA observations AL445, AC308 and AC101 (Fig.~\ref{fig:vla-all}). Although there is no obvious extended emission in this image, there are a number of resolved point sources. 

\begin{figure*}
 	\center{\includegraphics[width=0.5\textwidth,trim=0 90 0 0,angle=270]{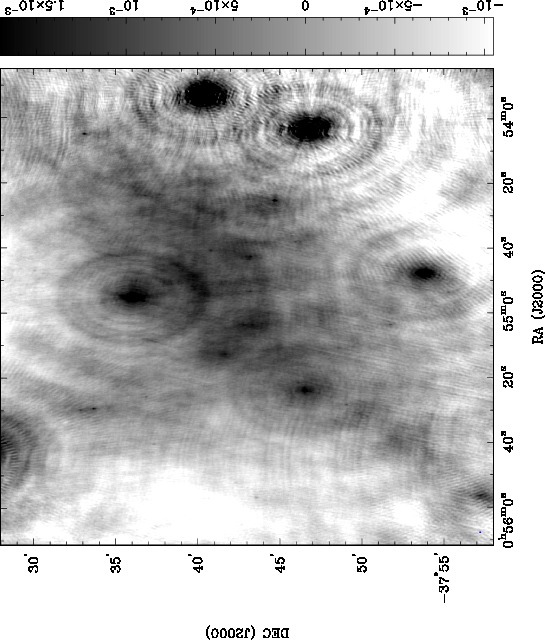}}
	\caption[All ATCA and VLA projects jointly deconvoled]{All ATCA and VLA projects radio-continuum total intensity image of \NGC. The synthesised beam, as represented by the blue ellipse in the lower left hand corner, is $11.8\arcsec \times 5.0$\arcsec\ and the r.m.s. noise is 0.13~mJy/beam. This image is in terms of Jy/Beam.  \label{fig:invert-all}}
\end{figure*}

\begin{figure*}
 	\center{\includegraphics[width=0.5\textwidth,trim=0 90 0 0,angle=270]{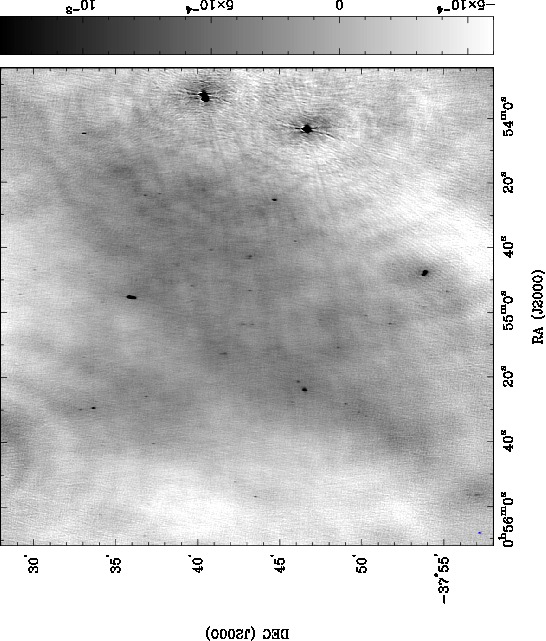}}
	\caption[All ATCA and VLA projects merged]{All ATCA and VLA projects radio-continuum total intensity image of NGC 300 produced using the \textsc{miriad} task \textsc{immerge}. The synthesised beam, as represented by the blue ellipse in the lower left hand corner, is $10.8\arcsec \times 4.8$\arcsec\ and the r.m.s. noise is 0.06~mJy/beam. This image is in terms of Jy/Beam.  \label{fig:immerge-all}}
\end{figure*}

In Fig.~\ref{fig:invert-all} we show the mosaic image produced with all five observations. Although there is a clear region of extended emission and a number of resolved point sources, this image suffers from significant side-lobe distortion. Finally, in Fig.~\ref{fig:immerge-all} we present the result of merging together each mosaic image. We note a well-defined region of extended emission in this image, with a total integrated flux density estimated to be 4.62$\pm$0.01~Jy at 20~cm. The details of all new images presented here, including the various combined images, are summarised in Table~\ref{table:2}. Images were also produced using a uniform weighing scheme, but were disregarded because of the absence of the extended emission structure. 
\begin{table}[h!]
\small
\center
\caption{The details of ATCA/VLA single and merged projects of \NGC{} mosaics at 20-cm. \label{table:2}}
\begin{tabular}{lcccccc}
\hline
\emph{ATCA/VLA}&\emph{Beam Size}&\emph{r.m.s.} & Figure\\
\emph{Project}& (arcsec) & (mJy/beam)\\
\hline
C1757  & 153.1$\times$72.5    & 0.30 & 2\\
\smallskip
C828\p0& 13.8$\times$10.1    & 0.07 & 3\\
AL445  & 7.0$\times$4.5       & 0.15 & 4\\
AC101  & 77.9$\times$36.7     & 0.30 & 5\\
\smallskip
AC308  & 77.8$\times$44.7     & 0.50 & 6\\
All ATCA  & 13.8$\times$10.1    & 0.07 & 7\\
All VLA  & 10.9$\times$4.8   & 0.09 & 8\\
All (invert) & 11.8$\times$5.0    &0.13 & 9\\
All (merged) & 10.8$\times$4.8    &0.06 & 10\\
\hline
\end{tabular}
\end{table}

\section{DISCUSSION}

\subsection{Discrete Sources Within The Field Of \NGC}

Our most sensitive and highest-resolution image is shown in Fig.~\ref{fig:immerge-all}. Using this image, a total of 72 radio sources above 3$\sigma$ (0.18~mJy/beam) were identified within the area defined by \citet{2004A&A...425..443P} (Table~\ref{table:srcs}). All sources identified in this study were (Gaussian) fitted without subtracting the background which changes across the field. While the uncertainty is dependent on the flux density of a source (with larger uncertainties for weaker sources) we estimate that the overall uncertainty in source flux density is less than 10\% (see \S4.2). 

A previous radio-continuum study of \NGC\ was able to identify a total of 47 sources at the wavelength of 20~cm (Payne et al. 2004; combined Cols. 5 and 6 from Table 4). These sources were found by combining two independent observations by the ATCA and the VLA. Eleven of these 47 sources could not be identified in our final image. This difference can be attributed to the method of identifying sources which the previous study adopted. For a source to be considered real, it only had to appear in one of the two radio-continuum images. Of the eleven sources which were found not to be in our final image, none could be identified in both 20~cm images which were produced in the previous studies. This potentially indicates that these eleven sources are not real as they can only be found in a single 20~cm radio-continuum image. Alternatively, they may represent some sort of transient or variable sources (e.g. QSOs) with very low flux densities of around 3$\sigma$. Indeed, we found some seven sources (Table~\ref{table:srcs}; difference between the new flux densities (Col.~4) and previous (Cols.~5 or 6)) with very different flux densities ($>$20\%) indicating a small but significant population of variable sources.

We also compared our radio-continuum catalogue with a list of 28 well-established optically identified \HII\ regions in \NGC\ \citep{2009ApJ...700..309B}. We note that 11 of our radio sources (sources \#~2, 11, 18, 21, 25, 28, 34, 39, 51, 52 and 54) coincide (to within 5\arcsec) with \HII\ regions from this list. Our source \#\,39 (ATCA\,J005451.7-373940) was previously classified as a SNR in \citet{2004A&A...425..443P}. 

\begin{figure*}
 	\center{\includegraphics[width=0.5\textwidth,trim= 0 120 0 120, angle=270]{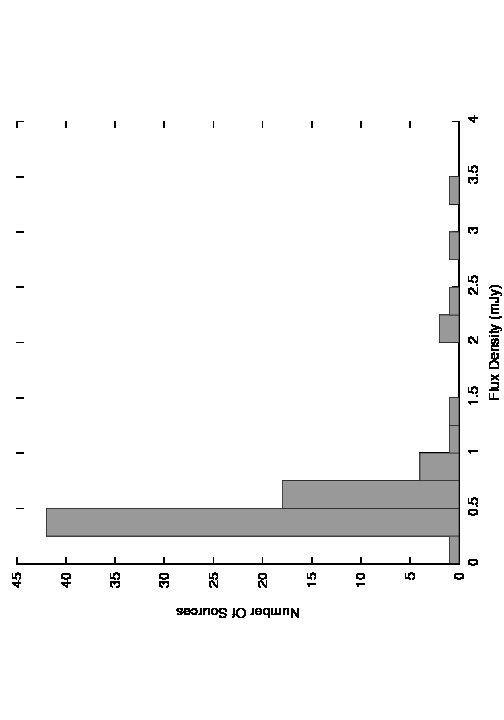}}
	\caption{A histogram of source flux density distribution within the field of \NGC. Source No.~42 from Table~\ref{table:srcs} is excluded from this graph.}
	\label{fig:ngc300-hist}
\end{figure*}

In Fig.~\ref{fig:ngc300-hist} we show the flux density distribution of all radio-continuum sources found in this study (Table~\ref{table:srcs}; Col.~3). As expected, the majority of sources (85\%) are within 10$\sigma$ flux density level.

\subsection{Comparison of Discrete Source Flux Densities}


While our flux densities are highly-correlated with those of \citet{2004A&A...425..443P}, they are systematically higher. Assuming no systematic offset between the two sets of flux 
densities, we find that our
flux densities (Col.~4 of Table 3) are about 20\%~ greater than those reported by \citet{2004A&A...425..443P} (ATCA observations in Col.~5 and VLA in Col.~6). The ATCA and VLA flux densities of \citet{2004A&A...425..443P} (Cols.~5 and 6) are in good internal agreement with ATCA/VLA=1.07$\pm$0.06 (standard error of mean for 33 sources in common). 
However, our new determinations of flux densities are higher than those of \citet{2004A&A...425..443P} by 19$\pm$5\%:
For the 8 sources from our study (Col.~4) which are stronger than 1~mJy, four sources (\#~9, 24, 42 and 62) had corresponding flux densities listed by \citet{2004A&A...425..443P} (Cols.~5 and 6). We found that from these four sources our flux densities are 23$\pm$9\%~ higher than the average of the ATCA and VLA observations; for all 35 sources in common with the ATCA observations (Col.~5), our flux densities are higher by 18$\pm$6\%; for the 33 sources in common with VLA observations (Col.~6), our flux densities are higher by 19$\pm$7\%.

The discrepancy between the flux density measurements is likely to be due to the extended continuum emission discussed below. In the earlier study, this component was resolved out
of the images, and therefore not measured in the point-source flux densities, whereas our image includes this emission, and our point-source flux densities also include it.

\begin{table*}
\tiny
\center
\caption{List of sources at 20~cm within the field of \NGC{}. Columns 1, 2, 3 and 4 are from this work. Columns 5 and 6 are partially extracted from \citet{2004A&A...425..443P}. Columns 7 and 8 correspond to the DSS2 (optical) and Spitzer (IR) identification.}
\begin{tabular}{@{}c l c c c c c c c l l}
\hline
\hline
 & 1                                      &         2                                                                   &   3   &   4  & 5     & 6       &  7 &  8  & 9 &10   \\
 & Identifier & $RA$ $(J2000)$                         & $Dec$ $(J2000)$                                                             & {\it S$_I$ }& ATCA & VLA   & Opt. & IR & Identifier & Source Type  \\ 
 & GFC2012&h \space m \space s \space\space\space &  \space \textdegree \space\space \arcmin \space \arcsec \space\space\space  & (mJy) & (mJy)& (mJy) &  ID     & ID & PFP2004 & \\ 
 \hline
1 & J005404.9-373305 & 00:54:04.89  & -37:33:05.02 & 2.34 &               &               &   &  & & \\
2 & J005416.6-373500 & 00:54:16.58  & -37:35:00.14 & 0.42 &               &               & y & Y & & \HII\\
3 & J005419.2-374322 & 00:54:19.23  & -37:43:22.58 & 0.36 &               &               &   &   & & \\
4 & J005421.2-373606 & 00:54:21.17  & -37:36:06.20 & 0.41 &               &               &   &   & &\\
\smallskip
5 & J005422.3-374022 & 00:54:22.25  & -37:40:22.49 & 0.38 &               &               & y &   & & \HII\\
6 & J005422.4-373613 & 00:54:22.43  & -37:36:13.62 & 0.78 & 1.37$\pm$0.04 &               &   &  &  J005422.5-373615 & \\
7 & J005423.5-373741 & 00:54:23.47  & -37:37:41.59 & 0.55 & 0.41$\pm$0.02 & 0.38$\pm$0.06 &   & Y &  J005423.4-373741 & \HII \\
8 & J005423.9-373649 & 00:54:23.91  & -37:36:49.47 & 0.68 & 0.64$\pm$0.02 & 0.58$\pm$0.06 &   &  &  J005423.8-373648 & \SNR \\
9 & J005425.2-374441 & 00:54:25.25  & -37:44:41.28 & 2.96 & 2.44$\pm$0.10 & 2.55$\pm$0.10 &   & Y &  J005425.2-374441 & \BKG \\
\smallskip
10 & J005427.3-374320 & 00:54:27.31 & -37:43:20.70 & 3.29 &               &               & y &   & & \HII\\
11 & J005428.8-374134 & 00:54:28.80 & -37:41:34.16 & 0.30 &               &               & y & Y & & \HII\\
12 & J005428.8-374351 & 00:54:28.80 & -37:43:51.31 & 0.39 &               &               &   &   & &\\
13 & J005430.7-374004 & 00:54:30.71 & -37:40:04.27 & 0.27 &               &               &   & Y & & \HII\\
14 & J005431.2-374556 & 00:54:31.23 & -37:45:56.95 & 0.51 & 0.41$\pm$0.03 & 0.33$\pm$0.04 &   & Y &  J005431.2-374554& \HII \\
\smallskip
15 & J005433.7-374158 & 00:54:33.69 & -37:41:58.04 & 0.17 &               &               &   &   & & \\
16 & J005434.0-373549 & 00:54:34.02 & -37:35:49.95 & 0.25 &               &               &   & Y & &\HII\\
17 & J005438.0-374557 & 00:54:37.96 & -37:45:57.65 & 0.66 & 0.65$\pm$0.03 & 0.66$\pm$0.06 &   & &  J005437.9-374559   & \BKG \\
18 & J005438.4-374147 & 00:54:38.35 & -37:41:47.51 & 0.30 & 0.27$\pm$0.02 & 0.34$\pm$0.06 & Y & Y &  J005438.1-374144 & \HII	\\
19 & J005438.4-374257 & 00:54:38.36 & -37:42:57.18 & 0.30 &               & 0.33$\pm$0.02 &   &  &  J005438.4-374240  & \SNR/\HII \\
\smallskip
20 & J005439.6-373541 & 00:54:39.57 & -37:35:41.55 & 0.46 & 0.42$\pm$0.02 & 0.43$\pm$0.10 &   & Y &  J005439.6-373543 	& \SNR \\
21 & J005440.7-374049 & 00:54:40.66 & -37:40:49.21 & 0.47 & 0.56$\pm$0.03 & 0.30$\pm$0.05 & Y & Y &  J005440.6-374049 & \HII	\\
22 & J005441.2-373349 & 00:54:41.19 & -37:33:49.77 & 0.34 & 0.29$\pm$0.02 & 0.33$\pm$0.05 &   & Y & J005441.0-373348 & \SNR \\
23 & J005442.3-374329 & 00:54:42.26 & -37:43:29.58 & 0.40 &               &               &   &   & & \\
24 & J005442.7-374313 & 00:54:42.66 & -37:43:13.09 & 1.07 & 0.63$\pm$0.07 & 0.69$\pm$0.06 & Y & Y &  J005442.7-374313 & \SNR/\HII \\
\smallskip
25 & J005443.4-374309 & 00:54:43.38 & -37:43:09.45 & 0.88 & 0.59$\pm$0.08 & 0.75$\pm$0.05 & Y & Y &  J005442.7-374313 	 & \HII\\
26 & J005443.8-373949 & 00:54:43.75 & -37:39:49.75 & 0.36 &               &               &   & &  & \\
27 & J005444.8-375226 & 00:54:44.79 & -37:52:26.04 & 0.39 &               &               &   &  & & \\
28 & J005445.1-373846 & 00:54:45.13 & -37:38:46.09 & 0.60 & 0.30$\pm$0.03 &               & Y & Y &  J005445.3-373847 & \HII \\
29 & J005445.1-375155 & 00:54:45.14 & -37:51:55.68 & 0.34 &               &               &   &   & & \\
\smallskip
30 & J005447.8-373324 & 00:54:47.83 & -37:33:24.06 & 0.29 & 0.31$\pm$0.03 & 0.37$\pm$0.07 &   & Y & J005448.0-373323 & \HII\\
31 & J005448.0-375352 & 00:54:47.97 & -37:53:52.24 & 2.02 &               &               &   &  & & \\
32 & J005448.8-375254 & 00:54:48.80 & -37:52:54.21 & 0.37 &               &               &   &  & & \\
33 & J005450.0-373616 & 00:54:49.99 & -37:36:16.96 & 0.37 &               &               &   &  & & \\
34 & J005450.2-373822 & 00:54:50.19 & -37:38:22.43 & 0.61 & 0.36$\pm$0.05 & 0.23$\pm$0.01 & Y & Y & J005450.3-373822 & \HII \\
\smallskip
35 & J005450.3-373849 & 00:54:50.26 & -37:38:49.74 & 0.55 & 0.24$\pm$0.02 &               & Y & Y & 	 J005450.3-373850 & \xrb \\
36 & J005450.3-375212 & 00:54:50.27 & -37:52:12.34 & 0.41 &               &               &   & Y & & \HII\\
37 & J005450.5-374129 & 00:54:50.52 & -37:41:29.11 & 0.33 & 0.28$\pm$0.02 & 0.30$\pm$0.03 &   & &  J005450.5-374123  & \BKG		\\
38 & J005451.0-373823 & 00:54:51.01 & -37:38:23.41 & 0.34 & 0.95$\pm$0.05 & 0.33$\pm$0.01 & Y & Y &  J005451.1-373826 & \SNR/\HII  \\
39 & J005451.8-373940 & 00:54:51.76 & -37:39:40.55 & 0.53 & 0.35$\pm$0.03 & 0.46$\pm$0.03 & Y & Y & J005451.7-373939 & \HII	 \\
\smallskip
40 & J005453.3-374312 & 00:54:53.29 & -37:43:12.57 & 0.62 & 0.69$\pm$0.04 & 0.66$\pm$0.06 & Y & Y &  J005453.3-374311 & \SNR/\HII \\
41 & J005453.5-375512 & 00:54:53.53 & -37:55:12.43 & 0.89 &               &               &   & Y & & \\
42 & J005455.3-373556 & 00:54:55.26 & -37:35:56.15 & 19.90& 17.88$\pm$1.31& 16.78$\pm$1.06&   & Y &  J005455.3-373557 & \BKG; Double	\\
43 & J005456.4-373939 & 00:54:56.41 & -37:39:39.75 & 0.31 & 0.39$\pm$0.06 & 0.68$\pm$0.06 & Y & Y & 	 J005456.3-373940 & \HII  \\
44 & J005456.8-373412 & 00:54:56.77 & -37:34:12.18 & 0.30 & 0.28$\pm$0.03 & 0.31$\pm$0.06 &   &   &  J005456.7-373413& \BKG \\
\smallskip
45 & J005458.4-375347 & 00:54:58.35 & -37:53:47.66 & 0.48 &               &               &   & Y && \HII\\
46 & J005458.4-375147 & 00:54:58.40 & -37:51:47.18 & 0.42 &               &               &   &   && \\
47 & J005500.9-373721 & 00:55:00.85 & -37:37:21.34 & 0.30 & 0.22$\pm$0.01 & 0.29$\pm$0.04 &   &  &  J005500.9-373720 & \BKG	 \\
48 & J005501.1-375016 & 00:55:01.00 & -37:50:16.79 & 0.47 &               &               &   & Y & & \HII\\
49 & J005502.3-374731 & 00:55:02.27 & -37:47:31.30 & 0.37 & 0.36$\pm$0.03 & 0.26$\pm$0.02 &   & &  J005502.2-374731  & \BKG	\\
\smallskip
50 & J005503.5-375144 & 00:55:03.54 & -37:51:44.38 & 0.82 &               &               &   & Y  & & \HII\\
51 & J005503.6-374248 & 00:55:03.56 & -37:42:48.17 & 0.46 & 0.34$\pm$0.02 & 0.41$\pm$0.03 & Y & Y & J005503.5-374246  & \HII \\
52 & J005503.7-374320 & 00:55:03.70 & -37:43:20.89 & 0.43 & 0.33$\pm$0.03 & 0.32$\pm$0.04 & Y & Y & J005503.6-374320 & \HII \\
53 & J005510.9-374834 & 00:55:10.85 & -37:48:34.24 & 0.62 & 0.52$\pm$0.04 & 0.30$\pm$0.05 & Y & Y &  J005510.8-374835 & \SNR/\HII \\
54 & J005512.7-374134 & 00:55:12.68 & -37:41:34.80 & 0.62 & 0.49$\pm$0.04 & 0.55$\pm$0.01 & Y & Y &  J005512.7-374140 & \HII\\
\smallskip
55 & J005512.8-374130 & 00:55:12.79 & -37:41:30.14 & 0.36 &               &               & y & Y & & \HII\\
56 & J005513.0-374417 & 00:55:13.02 & -37:44:17.28 & 0.28 &               &               & y &   & & \HII\\
57 & J005513.8-373305 & 00:55:13.83 & -37:33:05.01 & 0.41 &               &               &   &   & & \\
58 & J005515.3-374438 & 00:55:15.33 & -37:44:38.68 & 0.31 & 0.20$\pm$0.01 & 0.22$\pm$0.02 & Y & Y &  J005515.4-374439  & \SNR/\HII	 \\
59 & J005519.5-373605 & 00:55:19.54 & -37:36:05.00 & 0.25 &               &               &   & Y & & \HII\\
\smallskip
60 & J005521.3-374608 & 00:55:21.33 & -37:46:08.91 & 0.60 & 0.67$\pm$0.03 & 0.76$\pm$0.06 &   &   &  J005521.3-374609 & \BKG	 \\
61 & J005522.0-373635 & 00:55:22.00 & -37:36:35.10 & 0.45 &               &               &   & &  & \\
62 & J005524.0-374632 & 00:55:23.95 & -37:46:32.48 & 2.05 & 2.23$\pm$0.17 & 2.10$\pm$0.29 &   &  &  J005523.9-374632  & \BKG \\
63 & J005525.8-373652 & 00:55:25.82 & -37:36:52.57 & 0.61 & 0.63$\pm$0.05 & 0.55$\pm$0.05 &   &   &  J005525.8-373653& \BKG \\
64 & J005527.6-374546 & 00:55:27.60 & -37:45:46.76 & 0.37 & 0.36$\pm$0.03 & 0.30$\pm$0.01 &   &   &  J005527.6-374546& \BKG\\
\smallskip
65 & J005528.2-374902 & 00:55:28.23 & -37:49:02.18 & 0.75 & 0.96$\pm$0.05 & 0.77$\pm$0.06 &   &   &  J005528.2-374903 	& \SNR \\
66 & J005529.4-373339 & 00:55:29.36 & -37:33:39.25 & 1.37 &               &			    &   &   && \\
67 & J005529.8-373252 & 00:55:29.81 & -37:32:52.58 & 0.67 &               &			    &   &   && \\
68 & J005530.8-374951 & 00:55:30.80 & -37:49:51.05 & 0.53 &               &               & y & Y && \HII\\
69 & J005531.9-375016 & 00:55:31.92 & -37:50:16.83 & 0.42 &               &			    &   & Y && \HII\\
\smallskip
70 & J005540.4-373718 & 00:55:40.36 & -37:37:18.24 & 0.57 &               &               & y & Y && \HII\\
71 & J005552.2-374217 & 00:55:52.23 & -37:42:17.13 & 0.39 &               &			    &   &   && \\
72 & J005556.8-374331 & 00:55:56.76 & -37:43:31.14 & 0.73 &               &			    &   &   && \\

 \label{table:srcs}
\end{tabular}
\end{table*}


\subsection{Extended Radio-continuum Emission from \NGC}

\begin{figure*}
 	\center{\includegraphics[width=0.94\textwidth,trim= 0 120 0 120, angle=270]{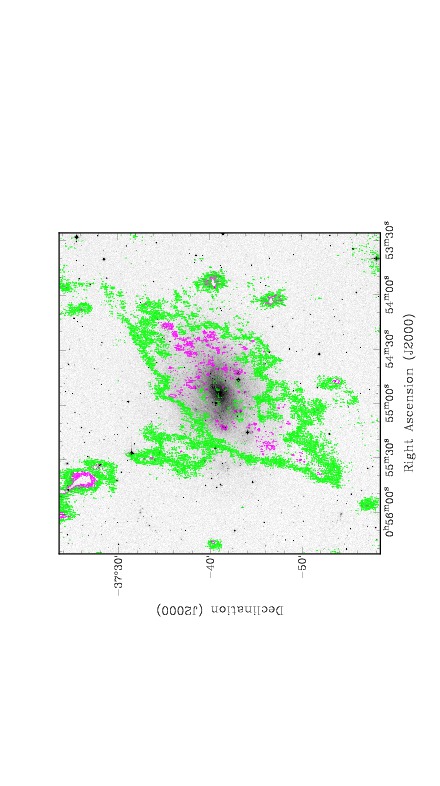}}
	\caption{An optical image of \NGC\ overlaid with our final radio-continuum image from Fig. \ref{fig:immerge-all}. The overlaid contours are 0.21 (3$\sigma$; green) and 0.49~mJy/beam (magenta). }
	\label{fig:ngc300-optical}
\end{figure*}

Fig.~\ref{fig:ngc300-optical} shows a DSS2 optical image of \NGC\ overlaid with our final radio-continuum contours from Fig. \ref{fig:immerge-all}. The structure of the radio emission from \NGC\ is significantly different from that of its optical emission.

Although the radio overlay aligns with a number of sources found in the optical image, there is a clear region of radio-continuum emission continuing past the apparent optical boundaries of \NGC\ and at a different position angle. The radio-continuum contours are consistent with the existence and positioning of the \HI\ outer disc spanning more than 1\textdegree\ (equivalent to about 35~kpc) across the sky \citep{2011MNRAS.410.2217W}. They suggested that there is a substantial change in the position angle between the inner and the outer disc, resulting in a twisted appearance of \NGC\ which is particularly obvious in the velocity field. One possible scenario for this twist is that the distortion and warping of the outer disc of \NGC\ was caused by tidal forces during a recent encounter with another galaxy, with nearby NGC\,45 as a potential candidate for a close encounter. 

While it is very difficult to precisely estimate the total flux density of \NGC\ extended radio-continuum emission at 20~cm we selected a tight region around 3$\sigma$ contour seen in Fig.~\ref{fig:ngc300-optical} and we summed all intensity inside this region to an integrated flux density of 4.687$\pm$0.005~Jy. This includes the extended emission to the north-east of the main disk.The total flux density of all point sources detected within this region is 0.067$\pm$0.005~Jy, so we estimate the flux density of the extended emission to be 4.62$\pm$0.01~Jy.

\subsection{Source Classification}

Thirty-five radio-continuum sources previously classified by \citet{2004A&A...425..443P} were detected in our study. These sources are classified into three groups: background sources (such as AGN), Supernova Remnants (SNRs) and \HII\ regions. To determine the true nature of the newly discovered sources from this survey, we used the previous classification scheme in addition to comparison with the available optical images such as DSS2. 

We found 8 new sources (marked as lower case {\it y} in Table~\ref{table:srcs}; Col.~7) that are in common with optical and radio images making a total of 24 sources in common between optical and radio surveys. These 8 newly detected optical sources are most likely to originate in \NGC\ as either SNRs or \HII\ regions. Because of their prominent optical extended emission, two sources (Table~\ref{table:srcs}; No.~40 and 53) which were previously classified as background AGN's are more likely to be intrinsic to \NGC. We also compared our new radio catalogue with existing \NGC\ IR Spitzer images \citep{2004ApJS..154..253H} at all frequencies (3.6, 4.5, 5.8, 8.0 24, 70 and 150 $\mu$m). We found 37 sources in common between radio and various IR frequencies. In total, 21 sources are common to all three surveys. Combining previous work and the new radio maps we classify 51 objects. From these 51 sources some 29 are most likely to be \HII\ regions, four are SNRs and one is an X-ray binary. We classify eleven sources as background sources (Table~\ref{table:srcs}; No.~9, 17, 37, 42, 44, 47, 49, 60, 62, 63 and 64 where 42 is a double source) and six sources being within \NGC\ (SNRs or \HII). Finally, the remaining 21 radio-continuum sources are still not classified.

We estimated the flux density of a bright Galactic SNR (such as Cas~A) at a distance of 1.9~Mpc, and compared this to our radio point source catalogue; we find that there are no
Cas~A analogues in \NGC. The luminous radio SNR source \#~62 (at about 2~mJy) is twice that of a Crab Nebula analogue (which would be about a 1~mJy source in \NGC). These results are similar to those in \citet{2000ApJ...544..780P}.

\section{CONCLUSION}

We present and discuss our new high-sensitivity and high-resolution radio-continuum images of \NGC\ at 20~cm. The new images were created from merging sensitive 20-cm mosaic radio surveys from ATCA, and from the VLA synthesis radio-telescope. We were able to dramatically improve both the sensitivity and resolution of the final radio-continuum image when compared to previous studies. This resulted in the identification of a number of previously unknown sources bringing the number of known radio sources within the area of \NGC\ to 72. We also detected a previously unseen region of \NGC's extended radio-continuum emission estimated to have an integrated flux density of 4.62$\pm$0.01~Jy. It was also demonstrated that a joint deconvolved image produced by both ATCA and VLA observations Fig. \ref{fig:invert-all}) was inferior to an image produced with an \textsc{immerge} (Fig.~\ref{fig:immerge-all}) approach. We believe that this difference is attributed to \textsc{miriad}'s inability to produce an accurate beam representation when datasets from multiple telescopes are jointly inverted. Existing sidelobes are exaggerated when cleaning the resulting dirty map because of the inaccurate beam.

\section*{Acknowledgments}
The National Radio Astronomy Observatory is a facility of the National Science Foundation operated under cooperative agreement by Associated Universities, Inc. The Australia Telescope Compact Array is part of the Australia Telescope National Facility which is funded by the Commonwealth of Australia for operation as a National Facility managed by CSIRO. This paper includes archived data obtained through the Australia Telescope Online Archive (http://atoa.atnf.csiro.au) and VLA.

We would also like to thank the anonymous referee, whose feedback greatly improved this paper.  

\bibliographystyle{spr-mp-nameyear-cnd}
\bibliography{ngc300paper}

\begin{thebibliography}{11}
\ifx \bisbn   \undefined \def \bisbn  #1{ISBN #1}\fi
\ifx \binits  \undefined \def \binits#1{#1} \fi
\ifx \bauthor  \undefined \def \bauthor#1{#1} \fi
\ifx \batitle  \undefined \def \batitle#1{#1} \fi
\ifx \bjtitle  \undefined \def \bjtitle#1{#1}\fi
\ifx \bvolume  \undefined \def \bvolume#1{\textbf{#1}}\fi
\ifx \byear  \undefined \def \byear#1{#1} \fi
\ifx \bissue  \undefined \def \bissue#1{#1} \fi
\ifx \bfpage  \undefined \def \bfpage#1{#1} \fi
\ifx \blpage  \undefined \def \blpage #1{#1} \fi
\ifx \burl  \undefined \def \burl#1{\textsf{#1}} \fi
\ifx \doiurl  \undefined \def \doiurl#1{\textsf{#1}} \fi
\ifx \betal  \undefined \def \betal{\textit{et al.}} \fi
\ifx \binstitute  \undefined \def \binstitute#1{#1} \fi
\ifx \binstitutionaled  \undefined \def \binstitutionaled#1{#1} \fi
\ifx \bctitle  \undefined \def \bctitle#1{#1} \fi
\ifx \beditor  \undefined \def \beditor#1{#1} \fi
\ifx \bpublisher  \undefined \def \bpublisher#1{#1} \fi
\ifx \bbtitle  \undefined \def \bbtitle#1{#1} \fi
\ifx \bedition  \undefined \def \bedition#1{#1} \fi
\ifx \bseriesno  \undefined \def \bseriesno#1{#1} \fi
\ifx \blocation  \undefined \def \blocation#1{#1} \fi
\ifx \bsertitle  \undefined \def \bsertitle#1{#1} \fi
\ifx \bsnm \undefined \def \bsnm#1{#1} \fi
\ifx \bsuffix \undefined \def \bsuffix#1{#1} \fi
\ifx \bparticle \undefined \def \bparticle#1{#1} \fi
\ifx \barticle \undefined \def \barticle#1{#1} \fi
\ifx \bconfdate \undefined \def \bconfdate #1{#1} \fi
\ifx \botherref \undefined \def \botherref #1{#1} \fi
\ifx \url \undefined \def \url#1{\textsf{#1}} \fi
\ifx \bchapter \undefined \def \bchapter#1{#1} \fi
\ifx \bbook \undefined \def \bbook#1{#1} \fi
\ifx \bcomment \undefined \def \bcomment#1{#1} \fi
\ifx \oauthor \undefined \def \oauthor#1{#1} \fi
\ifx \citeauthoryear \undefined \def \citeauthoryear#1{#1} \fi
\def \endbibitem {}
\ifx \bconflocation  \undefined \def \bconflocation#1{#1} \fi

\bibitem[\protect\citeauthoryear{{Bresolin} et~al.}{2009}]{2009ApJ...700..309B}
\begin{barticle}
\bauthor{\bsnm{{Bresolin}}, \binits{F.}},
\bauthor{\bsnm{{Gieren}}, \binits{W.}},
\bauthor{\bsnm{{Kudritzki}}, \binits{R.-P.}},
\bauthor{\bsnm{{Pietrzy{\'n}ski}}, \binits{G.}},
\bauthor{\bsnm{{Urbaneja}}, \binits{M.A.}},
\bauthor{\bsnm{{Carraro}}, \binits{G.}}:
\bjtitle{\apj}
\bvolume{700},
\bfpage{309}
(\byear{2009}).
doi:\doiurl{10.1088/0004-637X/700/1/309}
\end{barticle}
\endbibitem

\bibitem[\protect\citeauthoryear{Gooch}{2006}]{karma}
\begin{bbook}
\bauthor{\bsnm{Gooch}, \binits{R.}}:
\bbtitle{Karma Users Guide}.
\bpublisher{ATNF},
\blocation{Sydney}
(\byear{2006})
\end{bbook}
\endbibitem

\bibitem[\protect\citeauthoryear{Greisen}{2010}]{aips}
\begin{bbook}
\beditor{\bsnm{Greisen}, \binits{E.}} (ed.):
\bbtitle{Aips Cookbook}.
\bpublisher{The National Radio Astronomy Observatory},
\blocation{Charlottesville}
(\byear{2010})
\end{bbook}
\endbibitem

\bibitem[\protect\citeauthoryear{{Helou} et~al.}{2004}]{2004ApJS..154..253H}
\begin{barticle}
\bauthor{\bsnm{{Helou}}, \binits{G.}},
\bauthor{\bsnm{{Roussel}}, \binits{H.}},
\bauthor{\bsnm{{Appleton}}, \binits{P.}},
\bauthor{\bsnm{{Frayer}}, \binits{D.}},
\bauthor{\bsnm{{Stolovy}}, \binits{S.}},
\bauthor{\bsnm{{Storrie-Lombardi}}, \binits{L.}},
\bauthor{\bsnm{{Hurt}}, \binits{R.}},
\bauthor{\bsnm{{Lowrance}}, \binits{P.}},
\bauthor{\bsnm{{Makovoz}}, \binits{D.}},
\bauthor{\bsnm{{Masci}}, \binits{F.}},
\bauthor{\bsnm{{Surace}}, \binits{J.}},
\bauthor{\bsnm{{Gordon}}, \binits{K.D.}},
\bauthor{\bsnm{{Alonso-Herrero}}, \binits{A.}},
\bauthor{\bsnm{{Engelbracht}}, \binits{C.W.}},
\bauthor{\bsnm{{Misselt}}, \binits{K.}},
\bauthor{\bsnm{{Rieke}}, \binits{G.}},
\bauthor{\bsnm{{Rieke}}, \binits{M.}},
\bauthor{\bsnm{{Willner}}, \binits{S.P.}},
\bauthor{\bsnm{{Pahre}}, \binits{M.}},
\bauthor{\bsnm{{Ashby}}, \binits{M.L.N.}},
\bauthor{\bsnm{{Fazio}}, \binits{G.G.}},
\bauthor{\bsnm{{Smith}}, \binits{H.A.}}:
\bjtitle{\apjs}
\bvolume{154},
\bfpage{253}
(\byear{2004}).
doi:\doiurl{10.1086/422640}
\end{barticle}
\endbibitem

\bibitem[\protect\citeauthoryear{{Millar} et~al.}{2011}]{2011Ap&SS.332..221M}
\begin{barticle}
\bauthor{\bsnm{{Millar}}, \binits{W.C.}},
\bauthor{\bsnm{{White}}, \binits{G.L.}},
\bauthor{\bsnm{{Filipovi{\'c}}}, \binits{M.D.}},
\bauthor{\bsnm{{Payne}}, \binits{J.L.}},
\bauthor{\bsnm{{Crawford}}, \binits{E.J.}},
\bauthor{\bsnm{{Pannuti}}, \binits{T.G.}},
\bauthor{\bsnm{{Staggs}}, \binits{W.D.}}:
\bjtitle{\apss}
\bvolume{332},
\bfpage{221}
(\byear{2011}).
doi:\doiurl{10.1007/s10509-010-0556-y}
\end{barticle}
\endbibitem

\bibitem[\protect\citeauthoryear{{Pannuti} et~al.}{2000}]{2000ApJ...544..780P}
\begin{barticle}
\bauthor{\bsnm{{Pannuti}}, \binits{T.G.}},
\bauthor{\bsnm{{Duric}}, \binits{N.}},
\bauthor{\bsnm{{Lacey}}, \binits{C.K.}},
\bauthor{\bsnm{{Goss}}, \binits{W.M.}},
\bauthor{\bsnm{{Hoopes}}, \binits{C.G.}},
\bauthor{\bsnm{{Walterbos}}, \binits{R.A.M.}},
\bauthor{\bsnm{{Magnor}}, \binits{M.A.}}:
\bjtitle{\apj}
\bvolume{544},
\bfpage{780}
(\byear{2000}).
doi:\doiurl{10.1086/317238}
\end{barticle}
\endbibitem

\bibitem[\protect\citeauthoryear{{Payne} et~al.}{2004}]{2004A&A...425..443P}
\begin{barticle}
\bauthor{\bsnm{{Payne}}, \binits{J.L.}},
\bauthor{\bsnm{{Filipovi{\'c}}}, \binits{M.D.}},
\bauthor{\bsnm{{Pannuti}}, \binits{T.G.}},
\bauthor{\bsnm{{Jones}}, \binits{P.A.}},
\bauthor{\bsnm{{Duric}}, \binits{N.}},
\bauthor{\bsnm{{White}}, \binits{G.L.}},
\bauthor{\bsnm{{Carpano}}, \binits{S.}}:
\bjtitle{\aap}
\bvolume{425},
\bfpage{443}
(\byear{2004}).
doi:\doiurl{10.1051/0004-6361:20041058}
\end{barticle}
\endbibitem

\bibitem[\protect\citeauthoryear{{Rizzi} et~al.}{2006}]{2006ApJ...638..766R}
\begin{barticle}
\bauthor{\bsnm{{Rizzi}}, \binits{L.}},
\bauthor{\bsnm{{Bresolin}}, \binits{F.}},
\bauthor{\bsnm{{Kudritzki}}, \binits{R.}},
\bauthor{\bsnm{{Gieren}}, \binits{W.}},
\bauthor{\bsnm{{Pietrzy{\'n}ski}}, \binits{G.}}:
\bjtitle{\apj}
\bvolume{638},
\bfpage{766}
(\byear{2006}).
doi:\doiurl{10.1086/498705}
\end{barticle}
\endbibitem

\bibitem[\protect\citeauthoryear{Sault and Killeen}{2006}]{miriad}
\begin{bbook}
\bauthor{\bsnm{Sault}, \binits{R.}},
\bauthor{\bsnm{Killeen}, \binits{N.}}:
\bbtitle{Miriad Users Guide}.
\bpublisher{ATNF},
\blocation{Sydney}
(\byear{2006})
\end{bbook}
\endbibitem

\bibitem[\protect\citeauthoryear{{Steer} et~al.}{1984}]{1984A&A...137..159S}
\begin{barticle}
\bauthor{\bsnm{{Steer}}, \binits{D.G.}},
\bauthor{\bsnm{{Dewdney}}, \binits{P.E.}},
\bauthor{\bsnm{{Ito}}, \binits{M.R.}}:
\bjtitle{\aap}
\bvolume{137},
\bfpage{159}
(\byear{1984})
\end{barticle}
\endbibitem

\bibitem[\protect\citeauthoryear{{Westmeier}
  et~al.}{2011}]{2011MNRAS.410.2217W}
\begin{barticle}
\bauthor{\bsnm{{Westmeier}}, \binits{T.}},
\bauthor{\bsnm{{Braun}}, \binits{R.}},
\bauthor{\bsnm{{Koribalski}}, \binits{B.S.}}:
\bjtitle{\mnras}
\bvolume{410},
\bfpage{2217}
(\byear{2011}).
doi:\doiurl{10.1111/j.1365-2966.2010.17596.x}
\end{barticle}
\endbibitem

\end{thebibliography}
 
\end{document}